\documentclass[iop,revtex4]{emulateapj}
\usepackage{lineno}

\shorttitle{SOAR Adaptive Module (SAM)}
\shortauthors{Tokovinin et al.}

\begin{document}

\title{SOAR Adaptive Module (SAM): seeing improvement with a UV laser}

\author{Andrei Tokovinin,
Rolando Cantarutti,
Roberto Tighe,
Patricio Schurter, 
Manuel Martinez,  
Sandrine Thomas$^1$,
Nicole van der  Bliek, 
}

\email{atokovinin@ctio.noao.edu}

\affil{Cerro Tololo Inter-American Observatory, Casilla 603, La
  Serena, Chile}

\altaffiltext{1}{AURA/LSST 950 N. Cherry Ave, Tucson AZ 85719, USA}

\begin{abstract}
The  adaptive  module  of   the  4.1-m  SOAR  telescope,  SAM, corrects
ground-layer  turbulence using  a UV  laser guide  star.  It  has been
commissioned in 2013 and it is in  regular science operation since 2014. SAM
works with  the CCD imager covering  a $3'$ field or  with the speckle
camera.  It  operates routinely and stably,  delivering resolution in
the $I$ band equal to the free-atmosphere seeing. This paper describes
the SAM system as a whole, providing essential reference for its users
and technical  information of interest  to instrumentalists. Operation
of the instrument, its performance, and science projects done with SAM
so far are reviewed.
\end{abstract}



\section{Introduction}
\label{sec:intro}

SOAR  Adaptive  Module  (SAM)  is  a  facility  adaptive  optics  (AO)
instrument  at the  4.1-m Southern  Astrophysical  Research Telescope,
SOAR.   It improves  the  natural seeing  by  partial compensation  of
turbulence near the ground.  Although various aspects of SAM have been
covered  in the  literature,  there  has been no  single  reference with  a
comprehensive   description   of  the   entire   instrument  and   its
performance. The  present article fills this lacuna.   It is primarily
intended  for  SAM  users  and   may  be  of  interest  to  instrument
developers.  On the other hand, the format of a journal article is not
well suited  for technical details such as  full optical prescription,
software  algorithms, etc.;  this  paper is  {\it  not} an  instrument
reference manual.

From  the outset,  the SOAR  telescope  was designed  to deliver  high
angular  resolution  over  a  relatively narrow  field,  unlike  other
wide-angle 4-m  telescopes \citep{SOAR}.  It contains  a fast actuated
tertiary mirror (M3) for compensation of atmospheric tilts.  Extending
the  compensation  order using  the  AO  technology appeared  natural.
However,  SOAR  is  used  mostly  at visible  wavelengths,  where  the
standard AO compensates turbulence only over a narrow few-arcsec field
and needs a  bright star or a  powerful laser to do so.  In 2002, when
SOAR  was  completed,  the   AO  technology  enabled  good  turbulence
correction  only  in the  infrared  (IR).   However,  low-order IR  AO
systems  still provided {\it  partial} correction  in the  visible and
could improve the resolution by  a factor of $\sim 2$, as demonstrated
by the  PUEO instrument  \citep{PUEO}.  Such partial  correction could
benefit  a large range  of science  projects done  at SOAR  at visible
wavelengths.

The  idea  of  correcting   only  low-altitude  turbulence,  known  as
Ground-Layer  AO (GLAO),  was first  formulated by  \citet{Rigaut}. In
GLAO,  the  compensation  quality  is  traded  against  uniformity  of
correction over a wide field.   Quantitative analysis of GLAO was done
later by \citet{Tok04}, see also \citet{Gemini-GLAO}. Although several
natural guide stars (NGSs) or  laser guide stars (LGSs) located at the
perimeter of  the field of  view are optimal for  sensing ground-layer
turbulence,  even  a single  LGS  at  low  altitude is  an  acceptable
solution for GLAO. In such case, the so-called cone effect effectively
reduces sensitivity to high-altitude turbulence.

The concept  of SAM, first presented  by \citet{Tok03}, is  based on a
single  LGS created  by  Rayleigh scattering  of  an ultraviolet  (UV)
laser.  The  use of  a UV laser,  advocated by  \citet{Angel2000}, has
several  advantages:  efficient  Rayleigh scattering  proportional  to
$\lambda^{-4}$, easy separation of the laser light from longer science
wavelengths,  and   no  ocular  hazards,  making  the   beam  safe  to
airplanes. The LGS  is needed for a high  sky coverage; an alternative
wave-front  sensor  (WFS) using  several  NGSs  would  deliver a  much
inferior compensation. At that time, an  AO system with a UV laser was
being    built   for    the   2.5-m    telescope   at    Mt.    Wilson
\citep{Thompson2002},  while  a similar  AO  instrument  with a  green
Rayleigh laser  was developed at  the 4.2-m William  Hershel telescope
\citep{GLAS}.   For technical  reasons,  those two  projects have  not
produced any  science results, but  we learned from  their experience.
The GLAO system with a green Rayleigh laser at the 6.5-m MMT telescope
\citep{MMT} is not  used in regular observing programs.   On the other
hand,  the Robo-AO instrument  with a  UV laser  at the  Palomar 1.5-m
telescope \citep{RoboAO} is highly productive \citep[e.g.][]{Law2014}.

The  SAM instrument  design was  driven by  the science  goal (improve
seeing  over  a  moderate  field  of  view  at  optical  wavelengths),
availability  of suitable  technology  (e.g. UV  lasers  and fast  CCD
detectors), and  technological trade-offs. In the  development of SAM,
we tried  to use commercial  or otherwise proven  components, whenever
possible.   New and/or  critical  elements were  first prototyped  and
tested.  We  consulted with  the  AO team  at  the  MMT telescope  who
generously shared their experience.

The SAM system as a  whole is presented in \S~\ref{sec:overview}.  The
main AO  module is described in \S~\ref{sec:AOM},  the laser subsystem
is  covered in  \S~\ref{sec:LGS}.  \S~\ref{sec:SW}  describes  the SAM
software,  and  \S~\ref{sec:perf}  provides  information  on  the  SAM
performance.   Science operation  is covered  in  \S~\ref{sec:sci}. In
\S~\ref{sec:sum} we discuss the  place of SAM among other ground-based
and space facilities, its future instruments and upgrades.


\section{System architecture and overview}
\label{sec:overview}

\subsection{Evolution of the concept}
\label{sec:concept}

The choice  of the deformable  mirror (DM) has  a large impact  on the
instrument design. An adaptive secondary mirror, like in the MMT, would be
an ideal technical solution, but it was beyond our budget and, in  hindsight,
would have  substantially delayed the project.  Initially  we hoped to
use  a small  electrostatic  DM from  OKO-tech  and to  build a  small
instrument  around it,  but the  stroke of  this DM  turned out  to be
insufficient \citep{DM79}.  We selected  instead the bimorph DM BIM-60
made by CILAS.   The initial optical relay based  on refractive optics
was replaced by  a classical scheme with a  pair of off-axis parabolic
mirrors (OAPs) and a pupil  diameter of 50\,mm.  Then the AO-corrected
beam has the same focal ratio and pupil location as at the uncorrected
telescope focus,  allowing to  use with SAM  any instrument  built for
SOAR  (within  allowable  mass  and  space limits).   SAM  acts  as  a
seeing-improvement interface between  SOAR and its science instrument.
SAM  feeds  corrected  images  to  its  internal  CCD  detector,  SAMI
(4K$\times$4K  CCD) or  to  a  visitor instrument  (this  port is  now
occupied by the speckle camera).

\subsection{System layout}
\label{sec:sys}

\begin{figure}[ht]
\epsscale{1.0}
\plotone{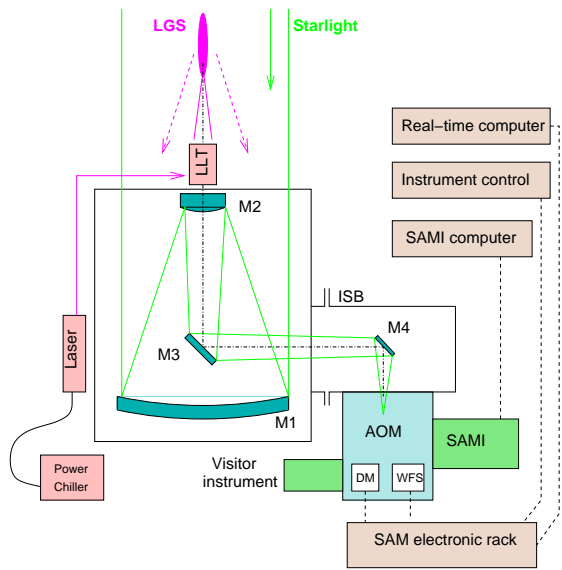}
\caption{\label{fig:all} Simplified block-diagram of the SAM
  system. The science light path is denoted by green lines, the laser
  light path is in pink. 
}
\end{figure}

Figure~\ref{fig:all}  presents a simplified  block-diagram of  the SAM
system.  Light  from celestial objects  reflected by the  SOAR primary
mirror M1  experiences three more reflections before  reaching the SAM
Adaptive-Optics Module (AOM),  attached to the instrument-selector box
(ISB) at the  Nasmyth focal station of SOAR.  During observations, SAM
rotates to compensate for  parallactic angle. The optical relay inside
AOM includes reflection from the DM. The corrected beam is sent either
to SAMI  or to  the visitor instrument, both mounted on  the opposite
sides of the AOM.

The LGS  is created by the  pulsed UV laser located in a box  on the
telescope truss  and also subjected  to variable gravity  depending on
the telescope elevation. The laser power supply and chiller are housed
in a thermally insulated rack attached  to the fork of the SOAR mount,
in fixed  gravity. The laser beam  is transported to  the laser launch
telescope (LLT)  located behind the  SOAR secondary mirror,  M2.  Each
pulse of the laser light propagates through the atmosphere and part of
it is scattered  back.  Photons scattered from the  distance of 7\,km
are selected  by the fast range-gate shutter  inside AOM, synchronized
with the  laser pulses.  The UV light  is analyzed  by the WFS,  and the
correction is sent to the DM in closed loop.  Other essential elements
of SAM shown in  Figure~\ref{fig:all} are its electronics, mostly housed
in a rack near the AOM, and its computers with software.

The  Rayleigh  scattering  preserves  polarization.  The  AOM  rotates
relative    to   the    telescope   and    the   laser,    hence   its
polarization-sensitive   WFS   cannot  be   adjusted   to  match   the
polarization of the laser. To  mitigate this effect, the laser beam is
propagated     circularly     polarized,     and     the     scattered
circularly-polarized beam is converted  back to linear polarization in
the  WFS.  Only the  oblique  reflection  from  M3
slightly affects the polarization tuning.

SAM  is   different  from  the majority of  AO  systems   in  several
ways. First, the tip and tilt (TT) compensation is provided by M3, the
actuated  tertiary  mirror  of  SOAR, upstream  from  the  instrument.
Second, TT is  sensed by two small guide probes  deployed in the input
(un-corrected) focal plane.  This avoids  the need to split photons in
wavelength between science and  guide channels.  The probes sample the
$5'$  square patrol  field outside  the $3'$  science field.   In this
design  the tilts  introduced by  the deformable  mirror (DM)  are not
sensed, but the DM control is tilt-free.  The UV light from the LGS is
focused well  behind the nominal  telescope focal plane,  meaning that
the light  path of LGS photons  inside SAM is very  different from the
starlight  path.  Nevertheless,  major non-common-path  errors between
the science and WFS paths were avoided.

Readers  who are  not interested  in the  SAM design  can skip  the
following material and go to Section~6.

\section{Adaptive Optics Module}
\label{sec:AOM}

In this Section, the AOM and its subsystems are covered. 

\subsection{Optical relay}
\label{sec:OAP}

\begin{figure}[ht]
\epsscale{1.0}
\plotone{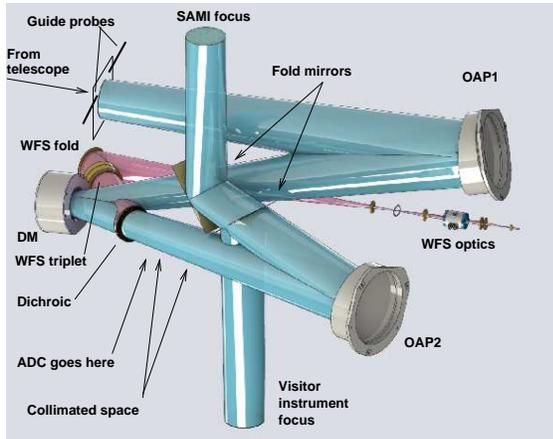}
\caption{\label{fig:OAP} AOM  optical layout. The  beam corresponds to
  the circular field of 3\arcmin (60\,mm) diameter.  }
\end{figure}

The   optical  layout   of  the   SAM   AOM   is  shown   in
Figure~\ref{fig:OAP},   and  the  optical   elements  are   listed  in
Table~\ref{tab:opt}. The beam, focused  by the telescope and reflected
by the M4 mirror inside the  ISB, propagates to the focal plane inside
AOM, where  guide probes  (GPs) are located,  and then hits  the first
OAP. Simulated  light source can be  inserted in the  path before OAP1
for  tests  and  DM   flattening.   The  collimated  beam  after  OAP1
propagates towards the DM, with an incidence angle of 14\fdg16.  After
the DM,  the UV light  is reflected by  the dielectric coating  of the
dichroic plate,  while longer wavelengths are  transmitted through the
dichroic towards OAP2 and  re-focused on the science instrument (after
the  fold mirrors).  The  SAMI fold  mirror is  installed permanently.
The  visitor fold  mirror  is  mounted on  a  translation stage;  when
inserted, it intercepts  the beam and directs it  to the visitor port.
The  atmospheric dispersion  corrector (ADC)  can be  inserted  in the
collimated  beam after  the  dichroic.  It  consists  of two  cemented
zero-deviation prisms, each made of  the BaK2 (angle 4\fdg45) and CaF2
(angle 5\fdg54) glasses \citep{ADC}.
Immediately  behind the ADC, the collimated  space is available
for  insertion of other  elements, for  example a  non-redundant pupil
mask or a Fabry-Perot etalon.

The UV photons  from the LGS at 7-km distance  are focused well behind
the science focal  plane, very close to the OAP1. The  LGS beam on the
DM is divergent.  Upon reflection  from the dichroic, it is focused by
the custom UV triplet lens, then  the beam is folded by a mirror.  The
triplet also works well for  the visible light, enabling SAM operation
with  the  NGS.  However,  this  mode  was  used only  during  initial
commissioning, and  SAM cannot be  reconfigured back to work  with NGS
without losing the LGS capability.

The  mirrors in the  AOM, including  the DM,  have a  protected silver
coating  with dielectric  layers enhancing  reflectivity at  the laser
wavelength 355\,nm.  The overall  transmission in the science path was
measured to be 0.90 at 633\,nm. During optical integration, the Strehl
ratio in  the science  channel at this  wavelength reached  0.60 after
alignment and with a carefully  flattened DM. This means that, despite
substantially  different light  path in  the WFS,  the non-common-path
errors are not significant (remember  that SAM optics does not need to
be diffraction-limited).

The optical  relay with two  OAPs has a  slightly curved field  with a
quadratic distortion. These effects, determined by the optical design,
were confirmed  by mapping the position  and focus of  the point light
source.  The curvature is small  and it can be safely neglected, while
the  distortion is  correctable  by post-processing  the images.   The
maximum lateral shift caused by the distortion is 0.6\,mm or 1\farcs8.

\subsection{Deformable mirror}
\label{sec:DM}

SAM uses  the bimorph  or ``curvature'' DM  BIM-60 made by  the French
company CILAS  for the 8-m  VLT telescope.  The  stroke of this  DM is
hence  more  than  sufficient   for  the  4.1-m  SOAR  aperture,  even
considering  the  reduced pupil  diameter  of  50\,mm  instead of  the
nominal 60\,mm.  The curvature radius with a maximum voltage of 400\,V
on all  electrodes is measured to  be 6.7\,m; a radius  of 30\,m would
suffice  to correct  a 1\arcsec  ~seeing in  SAM.  The  electrodes are
arranged in radial geometry with five rings.  Without any voltage, the
DM has a large static aberration, including defocus. About half of the
dynamic range  is used  to ``flatten'' the  DM. The  ``flat'' voltages
depend on ambient temperature  and instrument orientation, because the
thin DM is deformed by gravity as SAM rotates, creating mainly defocus
and trefoil. The gravity deformation  is compensated by applying to all
electrodes pre-calculated corrections varying  as sines and cosines of
the  rotation angle.   With those  corrections, SAM  can work  in open
loop, which  is useful  in its  normal operation when  the LGS  is not
needed or cannot be used for some reason.

\begin{table}[ht]
\center
\caption{Optical elements of the SAM AOM}
\label{tab:opt}
\begin{tabular}{l cc l}
\hline
Element & Diam. & Dist. next & Notes \\
        & (mm) & (mm)    &    \\
\hline
\multicolumn{4}{c}{\it Common path} \\
Focal plane & 60x60 & 810 & Science field \\
OAP1        & 150 & 890    & $F=810$\,mm \\
DM          & 60  & 150    & BIM-60 \\
%
\multicolumn{4}{c}{\it Science path} \\
Dichroic    & 80 & 32 & 614.7\,mm to OAP2 \\
ADC prism1  & 80 & 3 & Thickness 14.5mm \\ 
ADC prism2  & 80 & 550.6 & Thickness 14.5mm \\ 
OAP2        & 150 & 345.5/455.5 & To Visitor/SAMI \\
Visitor Fold  & 210x150 & 478.6 &  Visitor instrument  \\
SAMI Fold  & 210x150 & 368.6 &  SAMI  \\
Science focus & 60x60  & n/a & \\
\multicolumn{4}{c}{\it WFS path} \\
Dichroic    & 80 & 180 & \\ 
WFS triplet & 80 & 65 &  UV triplet \\
WFS fold &   100 & 616.3 &  Mirror \\
$\lambda/4$ plate &  25 & 6.6 & \\ 
F-adjuster lens  & 25 & 53.3  & Single lens \\
Field stop   & 0.55x0.55 & 61.3 & Square aperture \\
WFS collimator & 25 & 10.2 & Single lens \\
Polarizer-1 &   10 & 3 & FocTek polarizer \\ 
Pockels cell &   10 & (20) & QX1020  \\ 
Polarizer-2 &   10 &  10.2 & FocTek polarizer \\ 
Re-imager  L1 &  25 & 5.6 & \\ 
Re-imager  L2 &  25 & 43.3 & \\ 
S-H focal plane &  n/a & 12.2 \\
S-H coll. lens & 25  & 14.7 \\
Lenslet array &  2.2 & 6.5 & Pupil  1.989\,mm \\ 
CCD-39 &   2.2 & n/a & WFS detector \\ 
\hline
\end{tabular}
\end{table}

\subsection{Wavefront sensor}
\label{sec:WFS}

\begin{figure*}[ht]
\epsscale{1.0}
\plotone{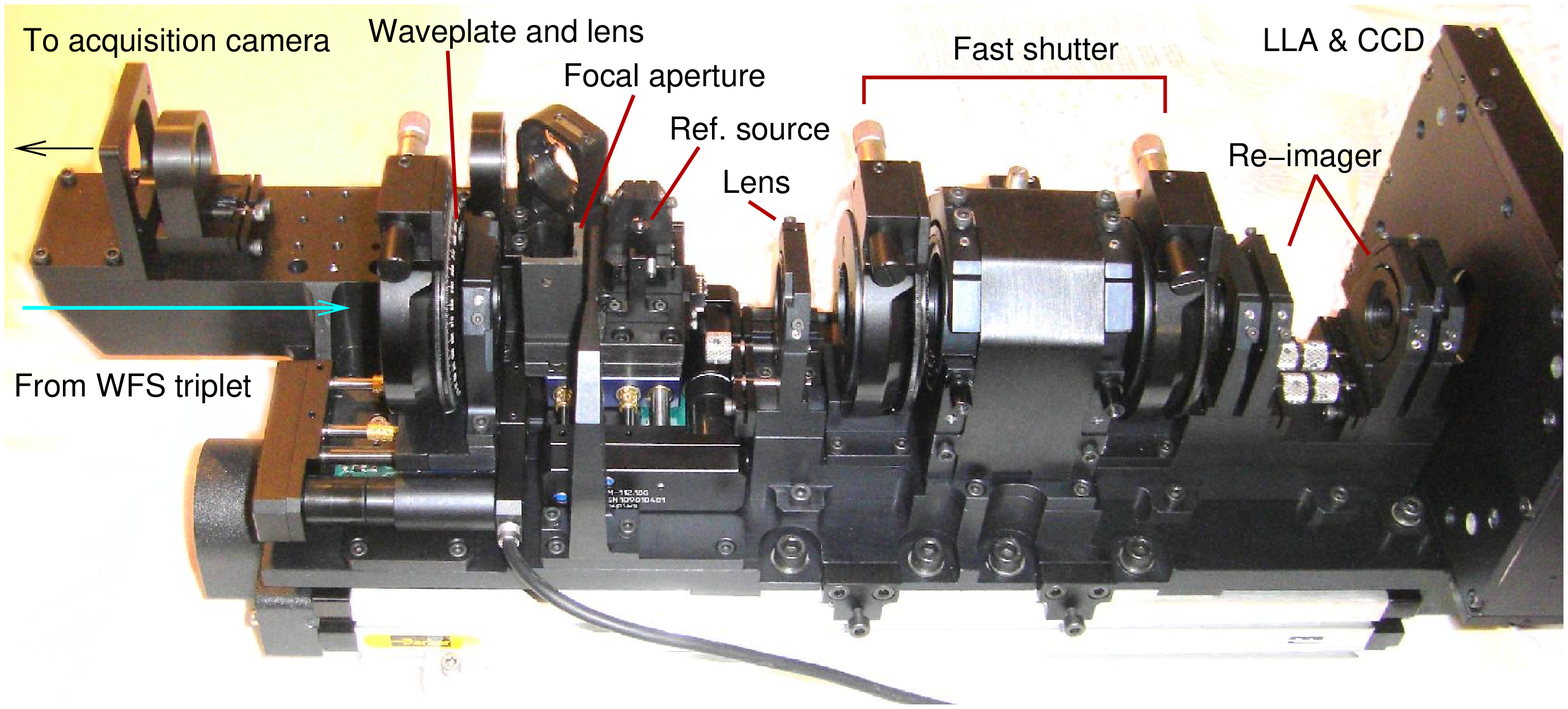}
\caption{\label{fig:LGS-WFS} LGS WFS.
}
\end{figure*}

SAM was first  tested on the sky in 2009 with  a simplified WFS, using
an NGS for turbulence sensing. The  UV dichroic was replaced then by a
neutral beamsplitter.  In the beginning  of 2011, when SAM was finally
commissioned in the NGS mode, the definitive LGS WFS was installed; it
is described  here. Both versions of  the WFS used  the same detector,
CCD-39, and the same lenslet array (LLA).

The LGS WFS optics (Figure~\ref{fig:LGS-WFS}) is tailored to match the
commercial  Pockels cell  QX1020 manufactured  by  Cleveland Crystals.
Constraints on the beam angular  divergence in the cell imposed by its
birefringent  material, KD*P,  and  the clear  aperture  of the  cell,
10\,mm, called  for a  $\sim$6-mm diameter beam  inside the  cell. So,
after the  field stop  at the WFS  focus the  beam is collimated  by a
fused-silica lens (good for  the monochromatic laser light). It passes
through  the first Glan-Taylor  polarizer, the  Pockels cell,  and the
second (crossed) polarizer.  Then two more singlet lenses re-shape the
beam and form  the pupil image of 1.99\,mm diameter on  the LLA, to get
10$\times$10  pupil sampling in  the Shack-Hartmann  WFS.  The  LLA is
placed directly  in front  of the CCD.   Its pitch,  0.192\,mm, equals
eight CCD pixels.  The LLA  focal length of 6.5\,mm (at 633\,nm) leads
to the pixel  scale of 0\farcs375, and the field  of view of 2\farcs8,
with 8 pixels per sub-aperture.

The optics  of SAM  allows the LGS  distance to be  set between  7\,km and
14\,km (it is fixed presently at  7\,km). When the WFS is refocused to
a different distance (it is mounted on a translation stage), the pupil
size on the  LLA changes slightly.  To compensate  for this variation,
another  weak singlet  lens  is placed  before  the focus  on a  small
translation stage.  Axial  motion of this focal adjuster  allows us to
keep  the constant  pupil diameter  at the  LLA. The  $\lambda/4$ wave
plate is placed together with  this lens to transform the elliptically
polarized laser beam into linear polarization.

The parameters  of the LLA APO-Q-P192-F5.75 were  chosen optimally for
the  NGS WFS,  while the  second LLA  with a  shorter focal  length of
3.17\,mm was procured  for the LGS WFS from  the same vendor, A$\mu$S.
Both LLAs are made of fused silica and have excellent optical quality.
However, we could  not use the short-focus LLA  as planned because the
distance  from  the CCD  surface  to  its  glass window  exceeded  the
specification by 1\,mm.  To match the  pixel scale and the size of the
LGS spots, we bin the CCD 2$\times$2, to 0\farcs75 per binned pixel.

The CCD-39 detector  has 80$\times$80 pixels of 24\,$\mu$m  size and four
output  amplifiers for  fast  readout. The  vendor, e2v, guaranteed  quantum
efficiency   of   $>$0.4  at   355\,nm.    The   detector  is   cooled
thermoelectrically to  about $-20^\circ$C.   CCD-39 works in  SAM with
the  SDSU-III  controller.   With  the 2$\times$2  binning,  the  loop
frequency is  478\,Hz and  the readout noise  is about  5.5 electrons.
 The digitized  signal is transmitted by the  optical fiber from
the controller to the  acquisition board inside the real-time computer
(RTC).   The   bias  pattern  is   not  very  stable,   requiring  its
re-calibration a few times per night, while slow drifts of the bias in
each quadrant  are tracked and  subtracted using signals in  the empty
corners of the image.

The duration of the range-gate pulse determines the vertical extent of
the  LGS and  hence  the  maximum spot  elongation  in the  peripheral
sub-apertures.   We normally use  the 150-m  range gate,  or 1\,$\mu$s
gate  pulses.   The amplitude  of  the  driving  pulses producing  the
$\lambda/2$ phase shift in the Pockels cell at the laser wavelength of
355\,nm is 2.7\,kV.  The KD*P crystal in the cell is piezoelectric, so
the driving pulses excite its  acoustic oscillations at 95\,kHz.  As a
result, the shutter does not close completely and some after-pulses or
``ringing''  appear.  The  1\,$\mu$s drive  pulses produce  a spurious
pulse  of  $\sim$10\% transmission  delayed  by  3.6\,$\mu$s from  the
opening edge of the gate.  The  weak ``tails'' of the LGS spots caused
by this ringing are truncated by the WFS entrance aperture, except the
innermost sub-apertures where they  pass through and slightly bias the
centroids.

The entrance  aperture of  the WFS is  carefully aligned  laterally to
avoid  ``spilling''  of   light  into  adjacent  sub-apertures.   This
aperture mask  is mounted  on a translation  stage which,  when moved,
replaces it by a UV source  used to measure the reference positions in
each sub-aperture.  At the same time the main beam is intercepted by a
diagonal mirror  that sends  it to the  acquisition camera.  We  use a
simple  GC650 CCD  detector from  Prosilica, with  a  Gigabit Ethernet
interface,  to acquire  ``live'' image  of  the un-gated  LGS.  It  is
strongly peaked and easy to center.  

\subsection{Tip-tilt guide probes}
\label{sec:GP}

Any LGS AO system needs a tip-tilt (TT) correction using natural guide
stars because the LGS does not provide valid TT signal. In the case of
GLAO, the requirements on residual TT errors are relaxed in comparison
to   the   classical  AO,   as   we  do   not   need   to  reach   the
diffraction-limited   resolution.   However,   the  sky   coverage  is
essential.  Our calculations indicated that  the faint limit of the TT
probes should be no less than $R=18$ mag for reaching the full sky coverage.

Atmospheric TT  errors increase with increasing  angular distance from
the TT guide star  (so-called tilt anisoplanatism).  The resulting PSF
non-uniformity was a  concern for SAM. We planned  to have three guide
probes (GPs)  picking stars  around the science  field of  view (FoV).
Further study showed that with  just two GPs the correction uniformity
is almost  as good,  so SAM has  two GPs,  each patrolling half  of the
field.  The  technical FoV for picking  the stars is  $5' \times 5'$
square.

\begin{figure}[ht]
\epsscale{1.0}
\plotone{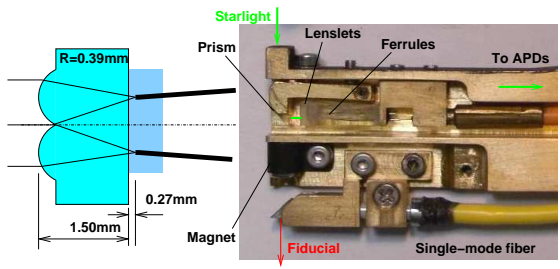}
\caption{\label{fig:GP} Optics  of the SAM guide probe  (left) and the
  actual device (right), with the light path in green overlayed.  }
\end{figure}

We  take  advantage  of the  SOAR  fast  tertiary  mirror for  the  TT
correction. The  GPs are placed  at the telescope focus  upstream from
the  optical relay.  To  reduce vignetting,  the width  of the  GPs is
minimized  to  6\,mm.  The  optical  scheme of  the  GPs  is shown  in
Figure~\ref{fig:GP}.    The  2$\times$2   array   of  0.5-mm   acrylic
micro-lenses, cut from  a larger sample, is optically  coupled to four
multimode fibers with a core  diameter of 100\,$\mu$m.  The SOAR pupil
image on  the fiber ends has  a diameter of  90\,$\mu$m, allowing some
tolerance for  alignment. Each fiber,  glued into a steel  ferrule, is
aligned with  respect to its micro-lens to  coincidence between images
of all  fibers formed by  the micro-lenses.  After alignment,  the lenslets
and ferrules are  potted by transparent epoxy to  form a solid module.
The ferrules  with fibers are slightly  tilted, so that  the source at
the center  is well within their acceptance  cone (numerical aperture)
which, together with the lenslet  size, define the field of view, 1\farcs5
per quadrant.  The four fibers  terminate by the FC connectors plugged
into the module of four photon-counting avalanche photo-diodes (APDs),
model SPCM-A Q4C from Perkin Elmer.  The measured optical transmission
is 0.7.  The acrylic lenslets do not transmit the UV light, so the probes
are not affected by the LGS even without any protective filters.

The  right-hand part  of  Figure~\ref{fig:GP} shows  the actual  guide
probe without  cover. The  light from the  telescope passes  through a
3-mm hole  and is deflected by  a 3-mm right-angle  prism before being
focused on  the lenslets  and injected into  the fibers. Tilts  of the
prism allow co-alignment of the optical axis with the direction of the
incoming light, i.e.  with the SOAR pupil.  The back side of the probe
facing SAM can  project a point source created  by a single-mode fiber
and  a  red  laser  diode.   This  fiducial  source  is  approximately
coincident with the star. It proved very useful for a number of tasks,
such  as mapping  the  guide-probe  coordinates on  the  CCD (to  ease
guide-star acquisition) and testing the optics of the AOM internally.

The APDs are cooled  thermoelectrically within their modules to reduce
the dark  count which  varies from 200  to 1500\,Hz, depending  on the
detector. The  high voltage (HV) enabling  the avalanche amplification
is turned  on only after positioning  the probes. To  protect the APDs
from damage  by over-light, we  installed a circuit that  switches off
the  HV as soon  as  the  current  exceeds some  threshold.   This
hardware protection has saved the APDs when a room light was turned on
accidentally during  SAM integration. Guide stars  (GSs) brighter than
$V \sim 11$ mag sometimes trigger the over-light protection.

Comparison of  the $V$  magnitudes of the  GS with  the actual
counting  rate  shows a  scatter  of about  1  mag,  mostly because  of
unreliable photometry in  the USNO-2 catalog.  The mean  flux $F$ (sum
of counting rates in the four quadrants) follows the expected trend
\begin{equation}
F {\rm [kel/s]} = 10^{ 0.4(V_0 -V) } ,
\label{eq:Flux}
\end{equation}
with  the zero points  $V_0$ of  19.9 and  19.7 mag  for GP1  and GP2,
respectively. These  zero points  match the estimated  sensitivity.  A
star of $V=18$ mag gives a flux of 5.75 kel/s (or 57 counts per 10\,ms
loop cycle)  in GP1. A Gaussian  star of 1\arcsec ~Full  Width at Half
Maximum, FWHM (rms  width 0\farcs42) will have a  centroid error on the
order of 0\farcs02 in the 10-Hz servo bandwidth. We do guide on $V=18$
mag stars when no better choice is left.

The 3\arcsec ~aperture  of the GPs is smaller  than the pointing error
of SOAR.  To acquire a  GS, we first  take the pointing  exposure with
SAMI and  determine the  offset by measuring  position of a  star with
known  coordinates in the  image.  The  GP is  then positioned  on the
selected GS  automatically, the HV on  the APDs is turned  on.  If the
star is detected,  the GP can be centered on  it manually; however, by
closing  the  TT loop  the  star  is  centered automatically.   Manual
centering of at least one GP is required if both GPs are used, so that
both  GPs work  near null.   If the  star is  not detected,  we  use a
software  tool  that modulates  the  tertiary  mirror circularly  with
increasing   radii  of   1\arcsec,  3\arcsec,   and   5\arcsec.   This
effectively enlarges  the capture zone  of the GPs.  If  the modulated
signal is detected,  the GP is moved towards  the star.  Otherwise, we
revise the pointing  offset or select another star  (some ``stars'' in
the USNO-2 catalog are galaxies not suitable for guiding).

\subsection{Turbulence simulator}
\label{sec:TurSim}

The AOM  contains a  built-in artificial UV  light source than  can be
injected in the optical path instead  of the LGS.  The focus and pupil
of  the  simulated  beam  mimic  the  actual LGS;  the  beam  can  be
artificially distorted, simulating the turbulence.  This device, called
TurSim, is essential for the  SAM operation; it is used for flattening
the DM and testing the health of the AO loop during daytime.

The  365-nm UV  light in  TurSim is  emitted by  the  photo-diode from
Nicia, with a  diffuser and a circular aperture  to emulate a 1\arcsec
~star.   Two  DC  motors  with  cam mechanisms  translate  the  source
laterally and serve  for centering its image in the  WFS.  The beam is
collimated by a custom UV  triplet lens, passes through the pupil mask
simulating central obscuration, and is re-imaged by another lens group
with  the   telescope's  focal  ratio,  at  a   fixed  axial  position
corresponding to  the LGS at 7\,km.   One or two phase  screens can be
inserted in the parallel beam near the pupil mask to simulate turbulence of
variable strength and  speed; they are driven by  regulated DC motors.
The phase screens, made  in-house \citep{TurSim}, do not match exactly
the  expected Kolmogorov spectrum,  but this  is  not critical  for
testing SAM.  TurSim has a deployable arm for projecting the simulated
LGS on-axis.  With a total of seven mechanisms, it is a rather complex
subsystem.  To transmit the TurSim  light through the WFS, its Pockels
cell is  triggered by internally  generated pulses rather than  by the
laser.

\subsection{AOM structure and mechanisms}
\label{sec:mech}

\begin{figure}[ht]
\epsscale{1.0}
\plotone{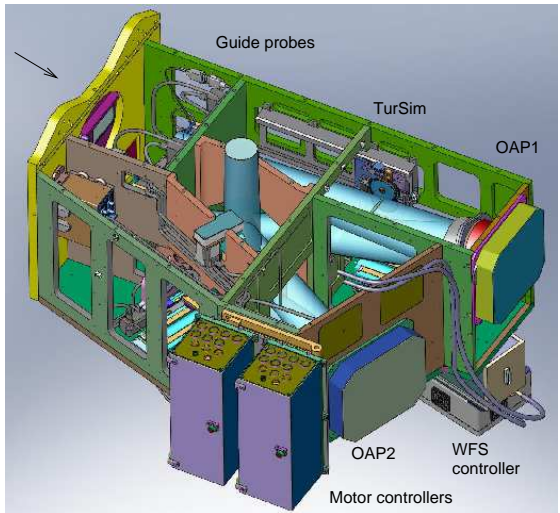}
\caption{\label{fig:mech}
Mechanical structure of SAM.
}
\end{figure}

All  elements of  SAM are  housed  in a  box-like aluminum  structure,
called bench  (Figure~\ref{fig:mech}).  It provides  stiffness against
gravity change in  one direction as SAM rotates  at the Nasmyth focus.
The total range  of flexure, measured by displacement  of the image of
the light source in the GP on the science CCD, is only $\pm$0\farcs35.
To prevent  image blurring by  flexure, avoid long exposures  near the
zenith associated with large rotation during the exposure.

The SAM  bench is made of  aluminum plates bolted  together. They have
large  rectangular openings  to access  the optics  and  mechanisms of
SAM. The covers are hermetized to prevent light and dust from entering
SAM. For  the same reason, the entrance opening of SAM is  covered by an
environmental shutter, opened only during observations.

The total mass of SAM (including SAMI) is 330\,kg. An instrument up to
70\,kg can  be mounted on the  visitor port. The  mechanical design of
SAM was complicated  because of the space restrictions  at the Nasmyth
focus. 

The AOM has 23 remotely controlled  motions (e.g.  6 in the guiders, 7
in TurSim, 3 in the ADC,  etc.). To reduce heat dissipation inside the
instrument,  all motor  controllers  are housed  in two  glycol-cooled
boxes attached  outside the  bench. However, the  APD modules  and the
driver of the  Pockels cell are located inside SAM, as  well as the WFS
detectors CCD-39 and GC650.

\subsection{SAMI, the CCD imager}
\label{sec:SAMI}

\begin{table}[ht]
\center
\caption{Parameters of SAMI}
\label{tab:SAMI}
\begin{tabular}{l  l}
\hline
Parameter & Value  \\
\hline
Format H$\times$V (pixels) &   4096$\times$4112  \\
Pixel size (arcsec) & 0.0455 \\
Field of view (arcmin) & 3.07 \\
Gain (el/ADU) & 2.1 \\
Readout noise (electrons) & 3.8 \\
Zero point   in SDSS $g'$ \& $i'$ (mag) & 24.4 \& 24.7 \\ 
\hline
\end{tabular}
\end{table}

The CCD imager, SAMI (Table~\ref{tab:SAMI}), uses the 4096$\times$4112
CCD231-84  detector with a  pixel size  of 15\,$\mu$m  manufactured by
e2v.   The chip has an astro  broad-band anti-reflection coating
providing a  quantum efficiency  of up to  0.90.  It has  an excellent
cosmetic quality  without any  major blemishes.  The  CCD is  read out
through four amplifiers using the  SDSU-III controller, as in the WFS.
The  full frame  without binning  is read  out in  about 8\,s,  with a
readout noise  of 3.8 electrons and  a gain of 2.1  electrons per ADU.
Most  of  the  time, the  imager  uses  a  2$\times$2 binning,  or  an
effective pixel scale of 91\,mas.  The CCD is housed in a dewar cooled
by liquid nitrogen and maintained at stabilized working temperature of
$-120^\circ$C.  A  blade shutter in  front of the CCD  allows exposure
time as short as 0.1\,s,  although longer exposures are recommended to
ensure  good  photometric uniformity.   The  filter  wheel is  mounted
before the shutter.

The  SAMI  software provides a standard  functionality: acquisition  of
single  or  multiple images  of  the  full frame  or  of  a region  of
interest.   The image  headers contain  information received  from the
telescope,  as well  as  the state  of  the SAM  instrument. The  SAMI
software can be driven externally,  e.g. by a script that synchronizes
exposures   with   stepping   of   the   Fabry-Perot   etalon   inside
SAM. Dithering can  be done by a script in  the SAM software. However,
recombination  of  images  taken  with large  dithers  requires  prior
correction of their distortion.

The  photometric zero point  of SAMI  in the  SDSS $g'$  band (stellar
magnitude  of  a  source  giving  a  flux of  1  ADU/s)  is  24.4  mag
\citep{Tok14}.  This corresponds to  the overall quantum efficiency of
0.46, which is quite high  considering light losses in the atmosphere,
telescope, SAM optics, and filter. The response of SAMI is higher than
that of SOI,  a simple optical imager at SOAR with  a focal reducer in
front of the CCD.

\section{Laser system}
\label{sec:LGS}

The LGS  is produced  by Rayleigh and  aerosol (Mie) scattering of  light at
355\,nm wavelength from the pulsed laser, projected on the sky through
a small telescope, LLT.  Main optical elements of the laser system are
listed in Table~\ref{tab:LGS} (CC stands for conic constant).

\begin{table}[ht]
\center
\caption{Optical elements of the SAM LGS}
\label{tab:LGS}
\begin{tabular}{l  l}
\hline
Element & Description  \\
\hline
Laser & Tripled Nd:YAG, 10W, 355\,nm \\
Beam expander & 8x magnification \\
LGS M4 mirror  & $D=25$\,mm \\
LLT fold mirror & UV mirror \\
Phase plate & UV  $\lambda/4$ \\
LLT M2 mirror    & Spherical  $D=15$\,mm, $R=$30\,mm \\
LLT M1 mirror    & $D=250$\,mm, $R=$840\,mm, CC=$-0.967$  \\
\hline
\end{tabular}
\end{table}

\subsection{The UV laser}
\label{sec:laser}

The  laser (model  Q301-HD from  JDSU) is  a  frequency-tripled  pulsed
Nd:YAG with  a nominal average power  of 10\,W.  It is  located on the
telescope  truss,   in  a  thermal  box   maintained  at  $+20^\circ$C
temperature and  flushed with dry  air.  In the real  operation (under
variable gravity and  at the altitude of Cerro  Pach\'on) the internal
power meter of the laser indicates  its output power in the range from
7.5\,W to 7.8\,W.  The beam quality measured in the laboratory is very
good,  $M^2  = 1.05$.   We  operate the  laser  at  its nominal  pulse
frequency of 10\,kHz. The pulse duration is about 34\,ns.
The JDSU company  fabricated hundreds of such lasers,   mostly for
photo-lithography, so  it is a  rugged industrial product. So  far, we
had no  problems with  the laser;  it works without any
servicing, apart from  the coolant refreshment every year  or so.

\subsection{Beam transfer optics}
\label{sec:BTO}

The narrow  beam emerging from the  laser is expanded  to the diameter
of $\sim$8\,mm (at $1/{\rm e}^2$ intensity level)  and directed to the
LLT  with  one reflection  from  the LGS-M4  mirror  at  the SOAR  top
ring. The beam  path to the LLT is enclosed in  a 1-inch aluminum tube
for safety  and dust protection. The  decision to expand  the beam was
taken to  reduce its intensity  on the path  to the LLT  (less coating
damage and  more tolerant  to dust  on the optics)  and to  have extra
flexibility  for adjusting  the diameter.  The  laser box  has a  fast
safety  shutter. During  alignment, a  mirror can  be inserted  in the
light path to substitute the UV beam with a green laser pointer.

Tilts of the  LGS-M4 mirror are controlled remotely  for centering the
beam on the  LLT.  Four UV photo-diodes around  the LLT primary mirror
and the 5th diode on-axis serve for the beam diagnostic.  The tilts of
the  LGS-M4  are regulated  to  center the  beam  by  balancing the  four
photo-diode signals approximately, while  the laser beam is propagated
in the dome during daytime.   Signals of these photo-diodes serve as a
useful diagnostic of normal LGS operation.

The  size  of  the  beam  on   the  LLT  mirror  is  adjusted  by  the
magnification and/or focus  of the beam expander in  the laser box. It
is  measured  by  the  signal  ratio of  the  peripheral  and  central
photo-diodes  and/or by  the shape  of the  UV beam  at the  LLT exit,
visualized by luminescence of  a standard white paper and photographed
(this beam  is not  very intense,  hence safe).  The  beam on  the LLT
mirror has FWHM diameter of 125\,mm.

\subsection{Laser Launch Telescope (LLT)}
\label{sec:LLT}

\begin{figure}[ht]
\epsscale{1.0}
\plotone{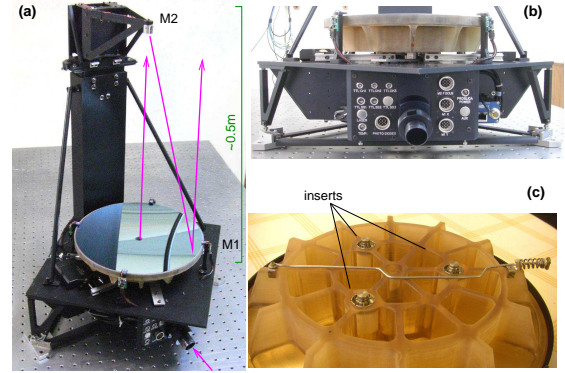}
\caption{\label{fig:LLT} The LLT. Photo on the left (a) shows the LLT
  without  enclosure, with the laser light path depicted in pink. The
  photo (b) is a side view from the input panel with connectors,
  showing how the primary mirror is suspended on the flexures. In (c)
  the back side of the light-weighted primary mirror is shown. The
  mirror diameter is 25\,cm. 
}
\end{figure}

The  design of the  LLT (Figure~\ref{fig:LLT})  is constrained  by the
small  space available  behind the  SOAR  M2. Its  25-cm aperture  was
chosen  to   minimize  the  LGS  spot  size,   balancing  between  the
diffraction  and the atmospheric  blur.  For  sodium lasers,  the LLTs
have typical diameters  of $\sim$0.5\,m, but for the  UV the optimum LLT
diameter is reduced in proportion to the wavelength. In hindsight, the
SAM LLT could be even  smaller because in reality the atmospheric blur
of the LGS dominates over diffraction.

The LLT optics  consists of the spherical secondary  mirror M2 and the
fast elliptical primary M1  (a Dall-Kirkham telescope).  Spherical M2 is
easier in fabrication and alignment,  but such optics has only a small
FoV, limited by the off-axis coma aberration. Although the LGS beam is
on-axis, the FoV  of $\sim 1'$ is needed  for co-alignment between LLT
and  SOAR,   compensating  the  flexure  of   both  telescopes.   This
requirement is  fulfilled by  mounting the M1 on  three flexure  rods that
point to its focus. Lateral motions of M1  are thus converted into
its rotations  around the focal point.  As M1 is actuated  to steer the
LLT beam on the sky, it  stays perpendicular to the beam, thus avoiding
the off-axis coma. The flexures also serve as a kinematic M1 mount. 

The LLT M1 was fabricated by S.~Potanin using a 5-cm thick Astro-Sitall (Zerodur
analogue) blank, light-weighted by  removing most of the material from
the back side and leaving  the rib structure.  Three  Invar inserts are
glued into the circular pucks  machined in the blank to connect mirror
with  the flexures.   Although  the thermal  expansion coefficients  of
Zerodur and  Invar are both about  $10^{-7}$, the 0.1-mm  layer of the
epoxy  glue around the  inserts produced  thermal deformation  of M1.
The problem was  fixed by replacing the solid inserts  with  flexible
equivalents that do not stress M1.

The M2  mirror is mounted above  M1 on a  column, attached to it  by a
light-weight  truss and  flexures. This  allows its  axial  motion for
focusing.  The distance  between M1 and M2 is fixed  by an Invar rod,
actuated for focusing by a motor  near M1.  In this way, the LLT focus
is  not very sensitive  to temperature  changes, despite  its aluminum
structure. Lateral  position of  M1  is defined by  two motors that
act on  its flexures laterally,  in a kinematic arrangement.   The LLT
internal structure is very stiff and light-weight.  It is connected to
SOAR  via three  mounting  points.  The  box-shaped  LLT enclosure  is
attached to the same  points independently of the main opto-mechanical
assembly.  The environmental shutter of  the enclosure is a thin steel
sheet that slides to the side when opened.

The 8-mm parallel  beam arriving at the LLT  after reflection from the
LGS-M4  is deflected  towards LGS-M2  by the  tertiary  mirror LGS-M3,
mounted  on the  piezo-electric tip-tilt  platform (S-330  from Physik
Instrumente) for  fast beam steering.   The platform is driven  by the
average spot centroids in the SAM  WFS and keeps the spots centered in
the WFS in closed loop. Remember that the SOAR M3 is actuated in TT to
stabilize the science  image; this also displaces the  LGS, so the LLT
TT servo has to compensate  for this disturbance and for the intrinsic
beam wander of  the LLT itself, to keep the  spots centered. The range
of  the  LLT  fast tilt  servo  is  $\pm$14\arcsec  ~on one  axis  and
$\pm$10\arcsec ~on the other  axis, reduced by oblique reflection from
the LGS-M3.   The LGS  can be steered  over a  larger range by  the M1
lateral actuators, while the piezo platform compensates only small and
fast residual tilts.

The LGS-M3 with  a dielectric UV coating is  almost transparent to the
visible light. Below  it, a small ``telescope'' is  located to capture
the collimated  light of a  bright star, for co-alignment  between LLT
and SOAR  and for testing the  LLT image quality.   It uses the  GC650 CCD
camera,  as in  the WFS  acquisition  channel.  This  device can  also
back-project  a parallel  beam  from  a red  laser  diode through  the
LLT. We mount a  20-cm flat mirror in front of the  LLT to reflect the
emerging red beam  back, and test the LLT  optics in auto-collimation.
This  functionality proved  critical  for {\it  in  situ} testing  and
alignment  of the  LLT. On  2015 September  16, the  strong earthquake
caused an  acceleration up to  $\sim$5g at the  SOAR M2.  The  LLT was
slightly misaligned,  but could be  re-aligned quickly by  tilting the
spherical  M2, using  the auto-collimation  mirror and  the  red diode
beam.

The size  of the LGS  spots in the  WFS often exceeded  2\arcsec, well
above the expected  seeing blur.  Partly it was  caused by the heat
dissipated  by the  LLT  electronics located  just  beneath it.   The
electronics was modified and is  now powered only when the LLT motors
are  actually  used.  As  a  result,  the  LGS spots  became  smaller.
However, they  are still enlarged  sporadically by internal  seeing in
the SOAR dome.

The  $\lambda/4$  wave  plate  located  above  the  LGS-M3  transforms
elliptical polarization  of the laser beam  into circular polarization
propagated  in the sky.   There were  several incidents  when insects,
attracted  by  the UV  light,  got inside  the  LLT  and were  burned,
damaging either the $\lambda/4$ plate  or the window of the laser box.
These elements were easily  replaced, without need for realignment. To
prevent further  damage, we covered  the LLT aperture with  a metallic
mesh (3-mm pitch and 0.1-mm thickness). It causes only a minor 
light loss of $\sim$12\% through blocking and diffraction.

\section{Computers and software}
\label{sec:SW}

\subsection{Real-time computer and software}
\label{sec:RTSoft}

The SAM real-time computer (RTC)  is a standard PC with an Intel$^{\rm
  R}$ Core Duo  processor E7500 running at 2.93\,GHz.   It works under
CentOS release 5.8 operational system (Linux kernel version 2.6.25 \#5
SMP) with a real-time patch RTAI~3.8.   The RTC is located in the SOAR
computer  room. It  contains the  acquisition  board of  the SDSU  CCD
controller receiving  the WFS signal and the  waveform generator board
for the M3 control.  Other interface  modules of the RTC are housed in
the  PXI chassis  in the  SAM  electronic rack.  The fiber  connection
between RTC  and PXI effectively  brings the computer bus  towards the
instrument.  The PXI  chassis must be powered when  the RTC is booted.
The  PXI  contains two  32-channel  digital-to-analog converters  that
generate  the  DM  drive  signals  and the  timer/counter  boards  for
acquisition of counts from the TT APD detectors.  Another timing board
generates the range-gate pulses of programmable delay and duration.

The  {\it  real-time  core}  (RTCORE)  is  at  the  heart  of  the  AO
control. Each control loop cycle is triggered by an interrupt from the
acquisition  board  when  the  fresh  CCD  frame  arrives.   The  spot
centroids are  computed using a  standard center-of-gravity algorithm.
Several  alternative  centroid   algorithms  such  as  correlation  or
weighted centroid \citep{Thomas06} are available and have been tested,
but  we found  that  with the  2$\times$2  binning and  the small  WFS
aperture the standard centroid gives the smallest noise. 

The  wave-front  slopes  are   the  differences  between  the  current
centroids and  the reference  spot positions. The  144 slopes  from 72
sub-apertures  are  multiplied   by  the  60$\times$144  reconstructor
matrix,  and   the  resulting  60  DM  drive   voltages  are  filtered
temporarily by the  loop controller (integrator with a  gain of 0.25).
The  reconstructor matrix  typically contains  38 to  40  system modes
(weak modes are rejected in the reconstructor calculation). To prevent
growth  of unseen modes,  a leak  of 0.01  is introduced.   A somewhat
faster second-order temporal controller (so-called Smith predictor) is
implemented,  but it  gives  no sizable  improvement  over the  simple
integrator.

When  using the  LGS,  the  standard AO  control  scheme is  modified.
Slopes  averaged  over  all  subapertures  are  subtracted  from  the
high-order control signals.  Those  average slopes instead are driving
the tip-tilt  platform in the  LLT to keep  the spots centered  in the
WFS.  The two control voltages for the LLT are generated together with
the 60 DM voltages.

Occasionally, the LGS  spots become blurred or contain  too few photons
because  of clouds.   Some spot  centroids then  remain indeterminate.
Those ``missing slopes'' do not  participate in the calculation of the
average tilt and are replaced by zeros in the high-order loop control.
 The AO loop is robust against  subapertures with missing information.

The  RTCORE communicates with  the non-real-time  applications written
mostly in LabView.  They include  the graphic user interface (GUI) and
various service modules. A bank of shared memory containing slopes and
DM voltages is used for the  analysis of the AO loop performance (e.g.
temporal power  spectra of  Zernike modes), estimation  of atmospheric
parameters and  residual aberrations.   Some of those  diagnostic data
are saved in the log file and in the image headers. Samples of AO data
(slopes, DM voltages, TT signals, and  WFS frames) can be saved by the
SAM operator for  further off-line analysis and archiving.

\subsection{Instrument control}
\label{sec:motion}

Apart from the real-time software, the SAM instrument is controlled by
four other  software modules running in the  instrument computer (IC),
also operated under Linux.  The IC  is housed in a cabinet on the SOAR
mount platform. All instrument control software is written in LabView;
each  module  has  its  own  GUI.  The software  is  accessed  by  VNC
connections either from the SOAR control room or remotely.

Two modules  (one for  AOM and one  for LGS)  take care of  the motion
control  and  other hardware.  The  motor  controllers, switches,  and
telemetry boards  of the  AOM are all  connected to the  serial RS-485
line acting as a bus and accessed by a communication board in the
IC.  The laser subsystems  are controlled  in a  similar way,  with an
addition of the direct serial communication with the laser.

Each  controlled motion  (e.g.   the  WFS focusing  stage)  has to  be
initialized  (search  for  the  zero-point mark)  after  powering  the
instrument.   Then  the  controllers  ``know'' the  current  mechanism
positions and can  move them using incremental encoders.   To speed up
the  start-up process and  to avoid  unnecessary motions,  all current
mechanism positions  are saved  by the software  and re-loaded  in the
controllers at the start, thus by-passing the initialization.  This is
particularly important  for the LLT  mechanisms that are  powered only
when actually used and hence  must ``wake up'' quickly.  The exception
is made for  the guiders: their six motors  are initialized after each
power-up.  This is done to avoid the risk of collision between the GPs
(their trajectories can overlap).   Collision is avoided by the motion
control software,  while an additional  hardware anti-collision sensor
is implemented as an extra precaution.

The  instrument  control software  (ICSOFT)  of  SAM  is a  high-level
LabView  application  used  to   operate  the  whole  instrument.   It
communicates with the lower-level  modules (motion control and RTSOFT)
and the telescope  through sockets.  In normal use,  SAM is controlled
only  through the  ICSOFT GUI  and there  is no  need to  access other
modules.   ICSOFT has  a  scripting capability  to  code sequences  of
actions, thus simplifying SAM operation. For example, search for guide
stars  by the  SOAR  M3 modulation  with  increasing amplitude,  their
detection and re-positioning of GP when the star is found is done by a
script activated by  a single button. There are  scripts for preparing
SAM for observations  and shutting it down at the end  of the night. A
software tool  for visualization of  the WFS acquisition images  and a
star catalog software for selecting  guide stars are run together with
the ICSOFT during SAM operation.

\subsection{Laser safety}
\label{sec:safety}

SAM operation is subject  to restrictions on laser propagation imposed
by  the US  Space  Command and  managed  by the  Laser Clearing  House
(LCH). Lists  of target coordinates are  sent to the  LCH several days
prior to  the observations, and  the authorized laser  propagation windows
for each target are received on the day of operation, or a day before.
These  files are  loaded  in the  software  module that  automatically
shutters  the laser  when the  telescope coordinates  differ  from the
target coordinates by more than 120\arcsec ~or when the propagation is
not authorized.  This software receives information from the telescope
control system.  Both the internal  laser shutter and the  fast safety
shutter in the laser box are closed. Typical LCH interrupts last a few
tens of seconds,  while their frequency varies from a  few per hour to
one or none per night, depending on the target location. Blanket laser
closures occasionally announced by  the LCH are introduced manually in
the software and then enacted automatically.

The SAM laser  does not present any hazards to  the airplanes.  Its UV
radiation  is  not visible  and  does  not  penetrate the  eye,  being
absorbed by  the cornea. As a  result, the safe  illumination level is
much higher  compared to the visible-light lasers.   An airplane going
through the SAM beam receives only  one laser pulse, as the next pulse
falls   at  a  different   non-overlapping  location;   the  resulting
illumination is  100 times  less than the  safe threshold for  eye and
skin established by the US safety standard ANSI Z136.6-200.

The narrow beam  of the SAM laser and even the  expanded 8-mm beam are
hazardous to  people.  Work with  those beams involves  standard laser
safety precautions  and can be  done only be qualified  personnel.  In
normal operation  these beams are  encapsulated inside boxes  and beam
conduits. The beam emerging from the LLT is sufficiently diluted to be
safe; moreover,  it is not  accessible from anywhere inside  the dome.
Consequently, the SAM  laser in normal operation does  not require any
safety  measures.   Nevertheless, a  luminous  sign  in the  telescope
control  room warns  when the  laser diodes  are energized.   A safety
circuit  and several emergency  stop buttons  are implemented  to shut
down the laser in abnormal situations.

\section{Performance of SAM}
\label{sec:perf}

\subsection{Requirements and error budget}
\label{sec:req}

The  basic requirements  follow  from  the intended  use  of SAM:  (i)
provide a  non-negligible improvement of the image  resolution, with a
FWHM of 2/3 compared to the uncorrected seeing, (ii) FWHM variation of
$<10$\% over the  field, and (iii) sky coverage of  no less than 90\%.
The  remaining  requirements  relate   to  the  instrument  use  (e.g.
one-person operation), operating conditions such as temperature range,
etc.   A large  number of  technical requirements  were  formulated to
guide the instrument design.

The quality of AO compensation is normally characterized by the Strehl
ratio, which translates to the  residual wavefront error.  In GLAO the
goal is to reduce the PSF size, so the Strehl ratio is not an adequate
metric. The  expected performance  of SAM was  first evaluated  for an
ideal  instrument,  considering  the  finite compensation  order,  the
geometry  of  the  LGS,  and the turbulence profiles  measured  at  Cerro
Pach\'on  \citep{TT06}.   The  PSF  is  computed  from  the  structure
function  of residual  wavefront errors  (RSF) which,  for  SAM, grows
almost  linearly  with the  distance.   The  corresponding  PSF has  a
sharper peak and stronger ``wings'' compared to the seeing profile; it
can  be well  approximated by  a Moffat  function with  $\beta  = 1.5$
\citep[][see  equation 1  below]{perf}.  

In  the  real  instrument,  the  errors  of  the  spot  centroids,  TT
residuals,  and  servo  lag  degrade the  performance  somewhat.   The
combined errors of  the SAM AO were required not  to degrade the ideal
resolution by more than 10\%, meaning that their relative contribution
to the RSF at the  characteristic baseline of 0.5\,m should not exceed
10\%.  In  this way, the error  budget and requirements  to various AO
subsystems   were  developed   \citep{perf}.    Unlike  the   standard
Strehl-based  approach  to  the  AO  design, it  constrains  only  the
difference between the real and ``perfect'' SAM and does not depend on
the fixed instrument parameteres and turbulence profile.

\subsection{Performance of the AO loop}
\label{sec:AO}

\begin{figure}[ht]
\epsscale{1.0}
\plotone{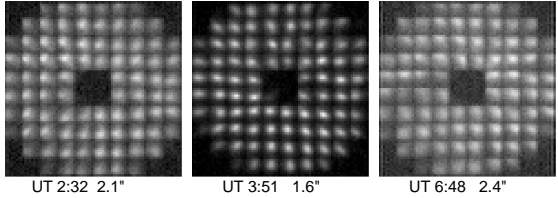}
\caption{\label{fig:spots}  LGS  spots in  the  SAM  WFS  in the  2013
  September run, illustrating the spot size variation from 1\farcs6 to
  2\farcs4, depending on the seeing.  }
\vspace*{0.1cm}
\end{figure}

When SAM was  first installed at SOAR in 2009, we  found that the axis
of the Nasmyth rotator did not point  to the center of the pupil. As a
result, rotation displaced  the pupil image on the  LLA. This has been
adjusted by tuning the angle  of the SOAR M3, so that rotation-induced
pupil  motion became  less  than  0.2 of  the  sub-aperture size.  The
diameter of  the pupil image on  the LLA was measured  by recording sky
through the  WFS, measuring flux  in each sub-aperture, and  fitting a
model  with three parameters:  pupil diameter  and its  lateral shifts.
The optimum  pupil radius is 5.1 sub-apertures,  the actually measured
one is  4.8 sub-apertures.  This causes  a loss of  flux in the peripheral
sub-apertures and  a mismatch between the  interaction matrix measured
with TurSim and the actual LGS.

With  a typical  range gate  of 0.15\,km  (maximum spot  elongation of
1\farcs25)  and the  loop frequency  of  478\,Hz the  return LGS  flux
varies  between  200  and  800  electrons per  sub-aperture  per  loop
cycle. The flux depends on the spot size (large spots are truncated by
the WFS aperture), atmospheric  conditions (more photons are scattered
by  aerosols or  less photons  due  to additional  absorption by  thin
clouds) and the rotator angle  (the polarization adjustment in the WFS
is not perfect).   Light losses in the up-link  and down-link LGS path
lead to  the estimated (and  measured) total efficiency of  5\%.  This
number, combined with  the lidar equation, implies the  return flux of
600  photons per  sub-aperture per  loop cycle,  as  actually observed
under favorable circumstances.

\begin{figure}[ht]
\epsscale{1.0}
\plotone{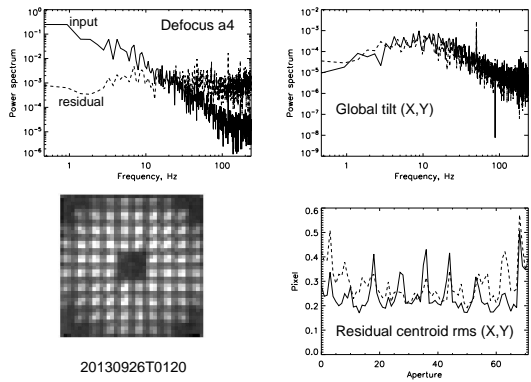}
\caption{\label{fig:loop}  Performance  of the  SAM  AO  loop on  2013
  September 26 at 1:20 UT.  Top left: temporal spectrum of the Zernike
  defocus coefficient  $a_4$ (residual in  dashed line, input  in full
  line, variance  1.2\,rad$^2$).  Top  right: temporal spectra  of the
  global tilt  in the WFS  in X (full  line) and Y (dashed  line), rms
  0.35 pixels.   Lower left: average  WFS spots (flux  160\,kel/s, spot
  size 1\farcs61).   Lower right:  residual centroid variance  in each
  sub-aperture (X in full line, Y  in dashed line) in WFS pixels, with
  the global tilt subtracted.  }
\end{figure}

The FWHM size of the LGS spots is measured by stacking many WFS frames
and fitting  Gaussians to several  spots around the pupil  center (the
elongation due  to the  finite range gate  is thus small).   It ranges
from  1\farcs3  to 2\farcs3,  depending  mostly  on  the local  seeing
(Figure~\ref{fig:spots}). The rms centroid  noise is normally between 0.15
and 0.25 pixels. Figure~\ref{fig:loop} gives an example of the AO loop
performance on a night with mediocre atmospheric conditions.

The TT  loop uses signals  of one or  two GPs to  control the
angles of SOAR M3.  The temporal controller accounts for the frequency
response  of  M3,  which  eventually  restricts the  TT  bandwidth  to
$\sim$10\,Hz. As the  TT probes are quad-cells, the  loop gain depends
on the  seeing.  With  a very good  seeing and  a gain too  large, the
15-Hz oscillation appears, so the nominal TT gain is set to avoid such
situations.  The  optical axis  of SOAR vibrates  with a  frequency of
50\,Hz and a typical rms amplitude between  20 and 50 mas. The cause of
this vibration is  not yet established, but it is  clearly seen in the
spectra of the TT signal.  This vibration is too fast to be compensated by
the TT loop.  Fortunately, its  contribution to the FWHM resolution is
only marginal  under normal use of  SAM and long  exposures. The 50-Hz
vibration is a very serious  factor in speckle interferometry, forcing to
use  exposures   shorter  than  10\,ms  and   affecting  the  limiting
magnitude.

\subsection{Delivered image quality}
\label{sec:DIQ}

\begin{figure}[ht]
\epsscale{1.0}
\plotone{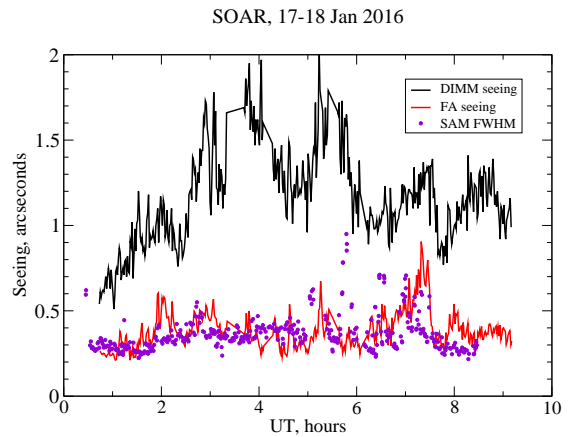}
\vspace*{0.5cm}
\caption{\label{fig:FA} Atmospheric conditions  and FWHM resolution of
  SAM  on  2016 January  17.   Series  of  short-exposure images  were
  recorded with HRCam in the $I$ band, co-added with re-centering, and
  approximated by  a Moffat function. The outlying  data points correspond
  to the data taken in open loop. The black and red curves correspond to the 
total and free-atmosphere (FA) seeing measured by the site monitor. }
\end{figure}

\begin{figure}[ht]
\epsscale{1.0}
\plotone{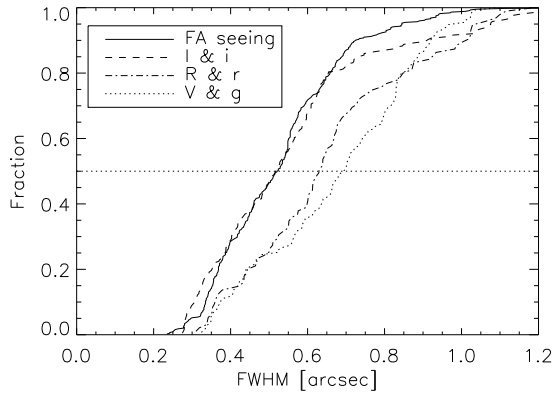}
\caption{Cumulative  histograms of  the  free-atmosphere seeing  (full
  line) and SAM  FWHM in various filters for  eight nights in January,
  February, and September 2013. The medians are 0\farcs52 for both the
  FA seeing  and the FWHM  in the $I$  or Sloan $i'$ bands,  while the
  median FWHM in  $R$ or $r'$ is 0\farcs63 and the  median FWHM in $V$
  or $g'$ is 0\farcs70.
\label{fig:diq}  
   } 
\end{figure}

\begin{figure*}[ht]
\epsscale{1.0}
\plotone{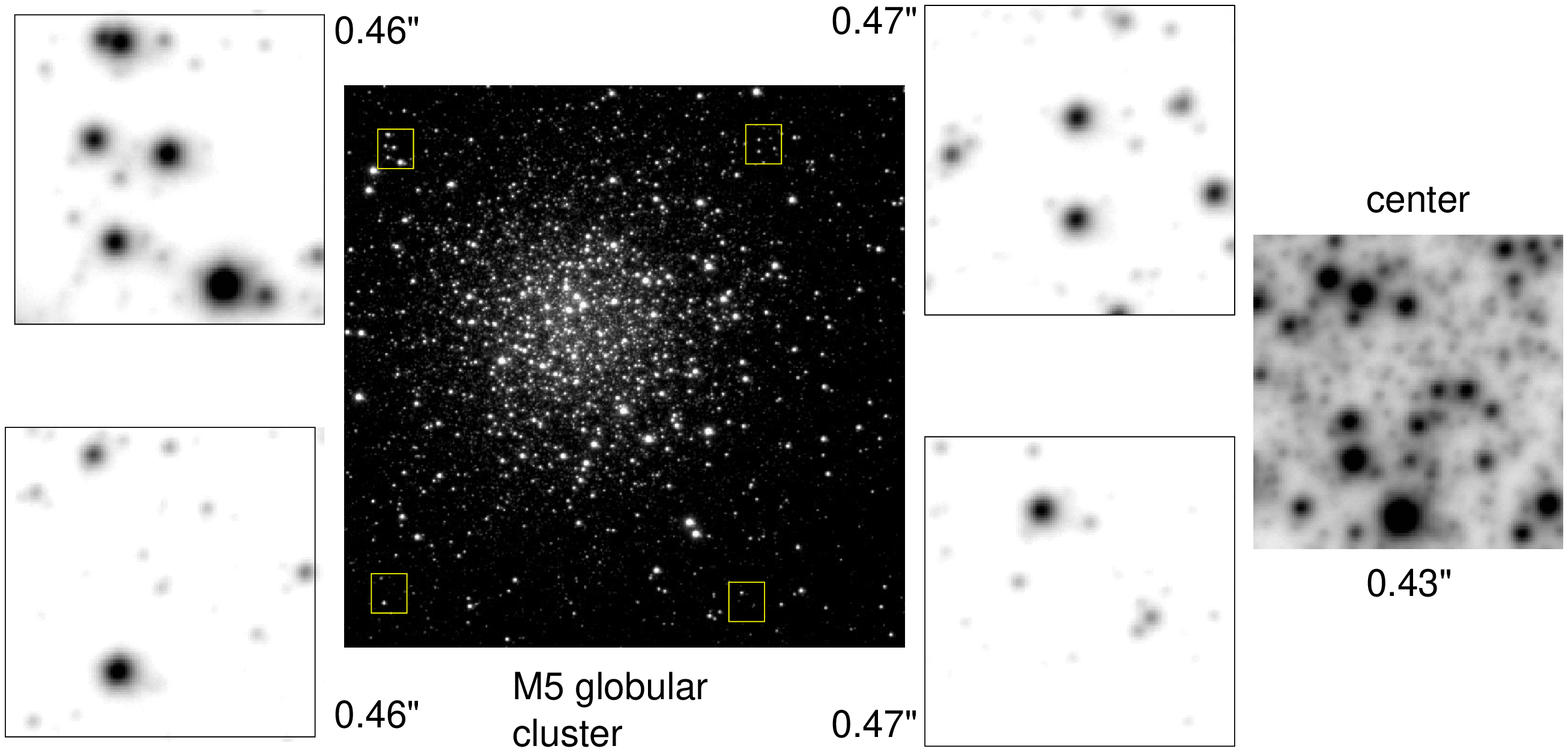}
\caption{Image of the  M5 globular  cluster taken  on 2013  March 2
  (six 10-s exposures in the $I$ filter, median-combined).  The square
  fragments of  12\arcsec ~size in the  corners and in  the center of
  the image demonstrate the stability  of the PSF over the entire field
  (the FWHM measured on several stars in each fragment is indicated).
\label{fig:M5}  
   } 
\vspace*{0.1cm}
\end{figure*}

The gain over  natural seeing brought by SAM  is quantified by several
metrics: by the FWHM resolution  and its uniformity over the field, by
the  PSF  profile and  its  variation,  and  by the  increased  energy
concentration.   The normalized  rotationally-symmetric PSF  $I(r)$ is
frequently approximated by the Moffat function
\begin{equation}
I(r) = [1 + (a r)^2]^{-\beta}
\label{eq:Moffat}
\end{equation}
with  two  parameters  $a$  and  $\beta$;  the  FWHM  is  $  (2  \sqrt
{2^{1/\beta} -1})/a$.   The case $\beta=1$ corresponds  to the Lorentz
profile,  whereas the  seeing profile  under Kolmogorov  turbulence is
well modeled by $\beta \approx 4.77$.

The  SAM PSF  under good  compensation corresponds  to  $\beta \approx
1.5$, in agreement  with its predicted shape (see \S~\ref{sec:req} above). Compared to
the seeing-limited PSF, it has  a sharper core and stronger wings. The
gain of SAM  depends on the metrics used and cannot  be expressed by a
single number.  Here we  quantify it by  the FWHM resolution.   If SAM
improves it by a factor of 2, the gain in the central PSF intensity is
about 2.8  times, while  the half-energy diameter  is reduced  only by
$\sim$1.5 times compared to the seeing.

The  FWHM resolution  at  long  wavelengths (e.g.   in  the $I$  band)
closely follows  the free-atmosphere (FA) seeing measured  by the MASS
instrument  (Figure~\ref{fig:FA}).   According  to  \citet{TT06},  the
median  FA  seeing  at   Cerro  Pach\'on  is  0\farcs40.   At  shorter
wavelengths, the  gain in resolution  over seeing is less  because the
compensation order  of SAM is  not high enough. Still,  the resolution
gain in the $R$ and  $V$ bands remains substantial.  However, when the
total seeing is dominated  by the high-altitude turbulence, SAM brings
no  improvement  because  it  does  not sense  this  high  turbulence.
Turbulence sensing  by SAM  decreases with distance  progressively and
turbulence at,  say, 0.5\,km  is compensated to  some extent,  but the
compensation is always  partial because of the finite  AO order, servo
lag, and noise.  So, the resulting SAM performance depends on a number
of factors. However, the simplistic  rule of thumb that SAM resolution
is almost  always close to the  FA seeing turned out  to be remarkably
good   in   practice,   as    demonstrated   by   plots   similar   to
Figure~\ref{fig:FA} that appear in \citep{AO4ELT3,Tok14}.

Figure~\ref{fig:diq}  shows cumulative  distributions of  the  FWHM in
several filters in eight nights during three SAM runs in 2013, under a
variety of turbulence conditions.  Each  point is a median FWHM of all
stellar sources in one  closed-loop exposure, not corrected to zenith.
The histogram of the FA sesing  at zenith measured by MASS, matched in
time to the SAM exposures, is plotted for comparison.  The FWHM in the
$I$ or $i'$ filters (214 estimates) equals the FA seeing in $\sim$80\%
of the  exposures, with the  same median of 0\farcs52.   The remaining
20\% are affected  by very poor total seeing  and, for deep exposures,
by semi-resolved distant galaxies which bias the median FWHM estimates
to larger  values.  On the other  hand, the 53  exposures in H$\alpha$
taken under  a median  FA seeing  of 0\farcs32 have  a median  FWHM of
0\farcs35.

The  uniformity of  correction  over  the $3'$  field  depends on  the
atmospheric conditions.  Both the  uncorrected high turbulence and the
well-corrected ground layer  do not cause PSF variation  in the field,
but the partially corrected turbulence  in the ``gray zone'' (at a few
hundred meters) leads to  the non-uniformity, with a better resolution
on-axis     \citep{Tok04}.      This     phenomenon    is     actually
observed,\footnote{See  Figure  21 in  the  SAM Commissioning  Report,
  \url{ http://www.ctio.noao.edu/soar/sites/default/files/SAM/archive/
    samrep.pdf}. }  but in general the  FWHM is very  uniform over the
field and rarely degrades by more  than a few percent towards the edge
(Figure~\ref{fig:M5}).    For   the   $I$-band   data   presented   in
Figure~\ref{fig:diq}, we  modeled the FWHM dependence  on the distance
from the  CCD center by  a linear function  and found that  the median
linear coefficient is zero, within errors.  The uniformity becomes even
better  at  shorter  wavelengths  where the  resolution  gain  becomes
smaller.  In  principle, the resolution should depend  on the distance
from the  TT star(s),  but this  effect has not  been studied  on the real
data.

\section{Science operation of SAM}
\label{sec:sci}

SAM has  been fully commissioned by  the end of  2013. Some scientific
results have  already been obtained  by that time  using commissioning
data \citep{Fraga2013}.  Since the  2013B semester, SAM was offered in
a  shared-risk mode,  and its  science verification  (SV)  program was
executed  in 2014. Regular  observing proposals  for SAM  are accepted
since  the 2014B  semester. Like  other SOAR  instruments, SAM  can be
operated remotely  from La Serena or  from other location  with a fast
enough network connection.

\subsection{Operational sequence}
\label{sec:oper}

SAM is  scheduled classically for specific  nights, preferably grouped
in runs (the instrument is switched  off between the runs but stays on
the  telescope).   Before  each  run,  the SAM  scientist  starts  the
software   and   makes    daytime   tests   following   the   standard
checklist. This procedure has been successful in detecting hardware or
software  problems,  leaving  time  for  their  fixes.   For  example,
failures of the line transformer in the high-voltage DM driver, failed
motor-control boards, or  a stuck motor shaft.  When  possible, SAM is
tested during engineering  nights a few days before  the science runs.
Fortunately, we  have not yet lost  any significant night  time to SAM
failures.

The filters  in SAMI are  installed by the telescope  operators during
daytime, as in other SOAR instruments.  SAMI can use either 4$\times$4
inch square  filters in  a wheel with  5 positions or  3$\times$3 inch
filters  in  another 7-slot  wheel.   The  limited  number of  filters
restricts  sometimes  the  number  of  science programs  that  can  be
executed with SAMI on any given night.

The  overhead time,  from the  start of  the telescope  slew on  a new
target to closing all four loops, can be as short as 5 minutes, with 7
minutes  being  typical.  Most of this  time is spent  for acquisition of
guide stars.  Owing to the small 3\arcsec ~field of the GPs, the exact
telescope pointing is determined by pre-imaging with SAMI; a star with
known  coordinates must  be identified  in this  image to  compute the
pointing offset.  {\it All} science  targets observed so far had guide
stars  down to  the $R=18$  limit, i.e.   the SAM  has a  complete sky
coverage.  Such a star gives $\sim$60 counts per 10-ms cycle of the TT
loop.

The SAM {\it observing tool} (OT)  is a custom IDL software that helps
formatting the target lists for the LCH, but also allows pre-selection of
guide stars and loading DSS2  images in the 5$\times$5 arcmin field to
assist in  field identification.  Guide  stars pre-selected in  the OT
can be fed to the SAM software and used alongside with the catalogs.

When the TT  loop is closed, the SAM operator  opens the laser shutter
and acquires  the LGS.  For LGS  centering, the un-gated  LGS image is
captured  by  the WFS  acquisition  camera.   The  LLT electronics  is
powered  on and  the LGS  is centered,  compensating  for differential
flexure between  the LLT  and SOAR optical  axes (typically  less than
10\arcsec).  Then the LGS tilt loop is closed, and the high-order loop
is closed as well (the LLT electronics is powered off automatically at
this time by the script).   The LGS acquisition is straightforward and
takes less than  a minute. If the telescope slew  was small, the laser
photons reach the WFS without re-centering  the LLT, in which case the two
LGS loops are closed within  a few seconds.  Experience shows that SAM
often remains operational  with a light cirrus clouds  because the LGS
is located below the clouds.

Time remaining until the next LCH interrupt is displayed in the ICSOFT. A
few  seconds  before the  interrupt,  the  LGS loops  are  opened
automatically, while  the science exposure  is normally paused  by the
observer. The SAM  operator closes the laser loops  manually after the
interrupt.

The first  target of the night  is usually used to  check the relative
focus between  SAM WFS  and SAMI.  The  focus might change  because of
temperature  or because  of  SAMI filters  of non-standard  thickness,
although it was found to be quite stable. To test the focus, two short
exposures are taken with closed  AO loop and WFS defocused by $\pm$4\,mm
from the nominal. The SAM PSF should be enlarged by the same amount
if the nominal focus is correct.  Otherwise, the correct focus setting
of the WFS  stage is found by linear  interpolation between bracketing
exposures.

SAM can operate  in open loop, without laser. In such  case, the DM is
flattened passively.  The LGS, if  available, still helps to focus the
telescope in  open loop by  measuring the focus  term and nulling  it by
adjusting the  telescope focus.  Open-loop operation  happens when the
image quality is  not critical or for targets that are  not in the LCH
list (e.g. photometric standards).

\subsection{Projects done with SAM}
\label{sec:prj}

During 2014,  SAM received only a  small number of  proposals, but its
use has steadily increased in  2015, making it the second most popular
instrument at SOAR.

\begin{figure*}[ht]
\epsscale{1.0}
\plotone{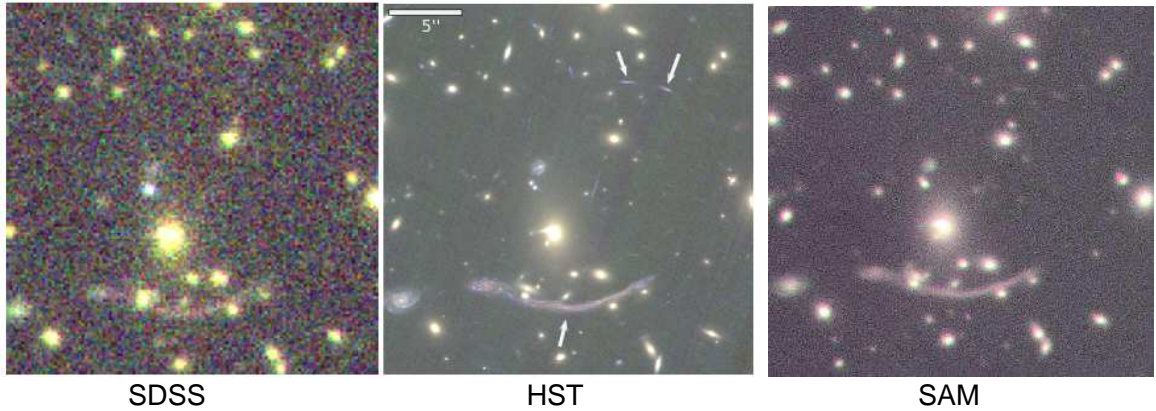}
\caption{Images  of  the  gravitational  arcs in  the  galaxy  cluster
  Abel~370 from SDSS,  HST, and SAM.  The SAM images  were taken on 2013
  September  29, 2013.  Five  5-minute  exposures in  the SDSS  $i'$ and  $r'$
  filters  each are  median-combined  and presented  as a  false-color
  image with a FWHM resolution of 0\farcs5.
\label{fig:arcs}
}
\end{figure*}

By concentrating the light of stars in a smaller number of pixels, SAM
reduces  photometric  errors  in  dense stellar  fields  dominated  by
confusion  \citep{Olsen}.  The  gain in  sensitivity in  such  case is
proportional to the central intensity  of the PSF. 

The first science paper  produced by SAM presented the color-magnitude
diagram of  the globular cluster NGC~6496  \citep{Fraga2013}. The FWHM
resolution  in  these  early   commissioning  data  of  2012  May  was
excellent, reaching  0\farcs25 in the  $I$ band.  However, it  was not
uniform over  the field,  showing an asymmetry  likely related  to the
wind  direction (the  SAM loop  worked then  with a  low  frequency of
233\,Hz). Despite  this, accurate  photometry was possible  by fitting
field-dependent  PSF with  DAOPHOT. Presently,  J.~Santos  (Brazil) is
conducting  a large  survey  of globular  clusters  in the  Magellanic
Clouds,  exploiting the  potential  of SAM  in  crowded fields;  eight
nights      in       the      2015B      semester       were      used
successfully.  \citet{Salinas2016} discovered  new  variable stars  in
globular clusters with SAM.

SAM can  help in  resolving close binary  stars, especially  the faint
ones beyond the reach of  speckle interferometry.  The survey of faint
low-mass  companions to  nearby stars  has revealed  one  new 0\farcs2
binary  \citep{Tok14}.   However,  for  binary stars  the  ``detection
power'' of SAM in terms of resolution and dynamic range is inferior to
the classical AO in the infrared or to speckle interferometry.

Combination of speckle interferometry  with laser-enhanced seeing is a
promising technique.  It  was used in 2015 May for  the survey of {\it
  Kepler-2}  field \citep{Kepler}.   SAM  worked with  the laser,  but
without guiding.  Series of short-exposure images were re-centered and
combined offline; the median FWHM resolution of the re-centered images
was 0\farcs43.   Binarity of targets as  faint as $I=15$  mag could be
probed down  to 0\farcs15 separation by the  speckle image processing.
This  technique was  used  again  in 2016  January  to discover  young
low-mass binaries in Orion, in  a project led by C.~Brice\~no.  The FA
seeing was better  than in May and the  FWHM resolution of re-centered
images reached 0\farcs25 (Figure~\ref{fig:FA}).

Several programs on emission-line  objects are planned using SAMI with
narrow-band  filters.   The  Herbig-Haro  jet HH~46/47  in  the  S[II]
narrow-band filter was observed with SAM on 2015 February 14, yielding
a FWHM resolution  of 0\farcs45.   Continued observations with  SAM 
use  a combination  of custom  narrow-band filters  to  probe physical
conditions in the jets.

The team of  A. Ardila observed with SAM  nearby galaxies NGC~1232 and
NGC~1482 in  the H$\alpha$ band  to study star-formation  regions with
high spatial resolution. The data are still being analyzed, while nice
images  resulting from this  effort were  generated with  a resolution
twice  better than  in  the image  of  NGC~1232  from  the VLT  
press release.

During three nights in 2015 SAM worked with a Fabry-Perot (F-P) etalon
inserted in its collimated beam  after the ADC, on a translation stage
\citep{FP}.   The existing F-P  etalon with  a spectral  resolution of
about 11\,000 just fits in the available space, with 1-mm gaps on both
sides. Selection of one F-P  order is done by the interference filters
in the SAMI  wheel.  This project is led by  C.~Mendez de Oliveira and
P.~Amram.   To  overcome  the  SAMI  readout  noise,  the  binning  of
4$\times$4 (pixels of 0\farcs18) was used.

Some science  verification programs intended to use  SAM for detecting
faint nebulae around  stars.  In this area, SAM  has no advantage over
other ground-based imagers;  it concentrates the light in  the core of
the PSF but does not reduce its wings.

SAM  is well suited  for high-resolution  imaging of  faint deep-space
objects.   Here  its  complete  sky  coverage and  a  moderately  wide
3\arcmin ~field  play an essential  role.  D.~Murphy used SAM  in 2013
for two nights to get a  very deep high-resolution image of a peculiar
galaxy  in  a  cluster,  apparently  affected  by  recent  interaction
\citep{Skidmark}.   Spectacular  images  of  thin  gravitational  arcs
produced by lensing distant galaxies were obtained with SAM during its
commissioning  (Figure~\ref{fig:arcs}).   No  proposals for  continuing
this  effort have  been  accepted  so far.   However,  the program  by
V.~Motta was started in 2015 December and continues in 2016.  Her team
took high-resolution images of  lensed quasar candidates revealed in a
wide-field survey.  This is an excellent example of the future 
use of  SAM, where it  will complement wide-angle surveys  by studying
interesting objects with higher spatial and/or spectral resolution.

Yet another promising application of SAM will be to study globular
clusters (GCs) in other galaxies. Even a small gain in resolution over
seeing might be critical here to distinguish GCs from stars. 
Results in this area are reported by \citet{Salinas2015}.

\section{Summary and outlook}
\label{sec:sum}

The crucial  role of the Hubble Space Telescope (HST) in astronomy,
despite its moderate 2.4-m size, demonstrates the need of high angular
resolution at optical wavelengths, so far not provided by ground-based
AO instruments. The resolution delivered by SAM is still much inferior
to the  resolution of the HST,  which continues to be  the facility of
choice.   The  advantages   of  SAM  compared  to  the   HST  are  its
accessibility  and the  larger  telescope aperture,  hence the  larger
number of collected photons.  When  very faint objects are observed in
narrow spectral bands, the minimum  size of the resolution element may
be set by the need to  collect enough photons.  In this situation, SAM
is preferable to the HST.  Its recent use with narrow-band filters and
Fabry-Perot nicely illustrates this aspect.

Under good seeing,  current AO instruments designed to  work in the IR
could  provide substantial  resolution gain  in  the red  part of  the
visible  spectrum, e.g. at  H$\alpha$. However,  their use  at optical
wavelengths is  not envisioned and sometimes is  impossible because of
dichroics in the science path or the lack of optical instruments. This
situation, presently to the advantage  of SAM, is slowly evolving. The
MUSE  instrument at  the VLT  \citep{MUSE} will  operate  with partial
seeing correction at visible  wavelengths over a 1\arcmin ~field using
four sodium lasers.  The ARGOS AO system at the LBT \citep{ARGOS} will
implement GLAO with green Rayleigh lasers and will work in the IR.

The potential  of SAM  will be fully  exploited when its  high spatial
resolution is  combined with spectroscopy.  This  is already happening
when SAM is  used with F-P. A multislit spectrometer  for SAM is being
developed by \citet{Robberto2016}.  This instrument, SAMOS, will use a
commercial digital micromirror device (DMD) as a software-configurable
multislit  mask.  The  spectral resolution  between 2000  and  8000 is
envisioned.   The SAMOS  science targets  are faint  extragalactic and
stellar objects  where the increased spatial resolution  brings a gain
in  sensitivity against  the sky  background, in  addition  to probing
small spatial details and reducing source confusion.

The SAM instrument designed more  than 10 years ago is technologically
obsolete.  Using  modern electron multiplication CCD  detectors with a
fast  gating, a higher-order  wavefront sensing  is feasible  with the
same laser power. If the DM is  replaced by a new device with a larger
actuator count,  the increased  compensation order will  improve image
correction  at shorter  wavelengths, in  the green  and  blue spectral
regions,  bringing  it  closer  to  the SAM  performance  in  the  $I$
band. Future upgrade of SAM along these lines will be decided based on
its popularity and availability of funds.

\acknowledgments Many people working at CTIO and SOAR have contributed
to the  creation of  SAM. Apart from  the authors,  the non-exhaustive
list  of  major  participants  in this  project  includes  B.~Gregory,
S.~Heathcote,   E.~Mondaca,    A.~Montan\'e,   W.    Naudy   Cort\'es,
F.~Delgado,  O.~Estay.    F.~Collao,  R.~Rivera,  D.~Sprayberry.   The
initial optical  design of  SAM by V.~Terebizh  and sharing of  the AO
code  by  Ch.~Keller helped  us  to start  the  SAM  project in  2002.
Consultations with M.~Hart and T.~Stalcup are gratefully acknowledged.
The efforts of  other experts in astronomical AO  who reviewed the SAM
project must not be forgotten.  S.~Potanin helped at a critical moment
by fabricating the  LLT mirror while major optical  vendors refused to
bid.   The  SOAR  crew  (E.~Serrano,  G.~Gomez,  G.~Dubo  and  others)
supported  SAM installation,  commissioning, and  operation. Referee's
comments helped to improve this article.




\begin{thebibliography}{}




\bibitem[Andersen et al.(2006)]{Gemini-GLAO}
Andersen, D. R., Stoesz, J., Morris, S. et al. 2006, PASP, 118, 1574


\bibitem[Angel \& Lloyd-Hart(2000)]{Angel2000}
Angel, J. R. P. \& Lloyd-Hart, M. 2000, Proc. SPIE, 4007, 270


\bibitem[Bacon et al.(2010)]{MUSE}
Bacon, R., Accardo, M., Adjali, L. et al. 2010, Proc. SPIE, 7735, 8


\bibitem[Baranec et al. (2012)]{RoboAO}
Baranec, C., Riddle, R., Ramaprakash, A. N. et al. 2012, 
Proc. SPIE, 8447, 4 (arXiv:1210.0532) 


\bibitem[Fraga et al.(2013)]{Fraga2013}  
Fraga,  L.,  Kunder,  A., \&  Tokovinin,  A. 2013, 
  AJ,  145, 165 

\bibitem[Hart et al.(2010)]{MMT}
Hart, M., Milton, N. M., Baranec, C. et al. 2010, Nature, 466, 727

\bibitem[Law et al.(2014)]{Law2014}
Law, N. M., Morton, T., Baranec, C. et al. 2014, ApJ, 791, 35


\bibitem[Mendez de Oliveira et al.(2016)]{FP}
Mendez de Oliveira, C., Amram, P., Quint, B. et al. 2016, MNRAS, submitted.

\bibitem[Murphy(2015)]{Skidmark}
Murphy, D. N. A. 2015, in IAU Symposium 309, p. 203

\bibitem[Olsen et al.(2003)]{Olsen}
Olsen, K. A. G., Blum, R. D., \& Rigaut, F. 2003, AJ, 126,  452


\bibitem[Rabien et al.(2014)]{ARGOS}
Rabien, S., Barl, L., Beckmann, U. et al. 2014, Proc. SPIE, 9148, 1


\bibitem[Rigaut et al.(1998)]{PUEO}
Rigaut, F., Salmon, D., Arsenault, R. et al. 1998, PASP, 110, 152

\bibitem[Rigaut(2002)]{Rigaut} Rigaut,  F., 2002, 
  in:   Beyond  Conventional   Adaptive   Optics,
  eds.  E.   Vernet,  R.  Ragazzoni,   S.  Esposito,  N.   Hubin,  ESO
  Conf. Workshop Proc. 58, ESO, Garching bei M\"unchen, p. 11 

\bibitem[Robberto et al.(2016)]{Robberto2016}
Robberto, M., Donahue, M,  Ninkov, Z. et al. 2016, Proc. SPIE, 9908, 99088V;
doi:10.1117/12.2233094


\bibitem[Rutten et al.(2006)]{GLAS}
Rutten, R., Blanken, M., McDermid, R. et al. 2006, NewAR, 49, 632


\bibitem[Salinas et al.(2015)]{Salinas2015}
Salinas, R., Alabi, A., Richtler, T. \& Lane, R. R.
2015, A\&A,  577, 59

\bibitem[Salinas et al.(2016)]{Salinas2016}
Salinas, R., Contreras Ramos, R., Strader, J. et al. 2016, AJ, accepted
(arXiv:1605.06517).

\bibitem[Schmitt et al.(2016)]{Kepler}
Schmitt, J. R., Tokovinin, A., Wang, Ji et al. 2016, AJ, 151, 159
(ArXiv:1603.06945S).

\bibitem[Sebring et al.(2002)]{SOAR}
 Sebring, T. A., Krabbendam, V. L.,  \& 
Heathcote, S. 2002, Proc. SPIE, 4837, 71

\bibitem[Thomas(2004)]{TurSim}
Thomas, S. 2004, Proc. SPIE, 5490, 766

\bibitem[Thomas et al.(2006)]{Thomas06}
Thomas, S., Fusco, T., Tokovinin, A. et al. 2006,   MNRAS,  371, 323

\bibitem[Thompson \& Teare(2002)]{Thompson2002}
 Thompson, L.  A. \& Teare, S.  W.  2002, PASP, 114, 1029

\bibitem[Tighe et al.(2016)]{ADC}
Tighe, R., Tokovinin, A., Schurter, P. et al. 2016, Proc. SPIE, 9908, 99803B;
doi:10.1117/12.2233681

\bibitem[Tokovinin(2004)]{Tok04}
Tokovinin, A., 2004,  PASP,  116,  941 


\bibitem[Tokovinin et al.(2003)]{Tok03} 
Tokovinin, A.  Gregory, B. \& Schwarz, H., 2003, 
 Proc. SPIE 4839, 673


\bibitem[Tokovinin et al.(2004a)]{SPIE04}
 Tokovinin, A.,  Thomas, S.,  Gregory, B. et al.
2004a, Proc. SPIE 5490, 870

\bibitem[Tokovinin et al.(2004b)]{DM79}
Tokovinin, A., Thomas, S., \& Vdovin, G. 
2004b,  Proc. SPIE,  5490,  580

\bibitem[Tokovinin \& Travouillon(2006)]{TT06}
Tokovinin, A. \&  Travouillon, T., 2006, MNRAS, 365,  1235

\bibitem[Tokovinin(2008)]{perf} 
Tokovinin, A., 2008,  Proc. SPIE 7015,  77


\bibitem[Tokovinin et al.(2008)]{SPIE08}
Tokovinin, A., Tighe, R., Schurter, P. et al. 2008,  Proc. SPIE 7015,  157



\bibitem[Tokovinin et al.(2010)]{SPIE10}
Tokovinin, A., Tighe, R., Schurter, P. et al. 2010,  Proc. SPIE, 7736,  132

\bibitem[Tokovinin et al.(2012)]{SPIE12}
Tokovinin, A., Tighe, R., Schurter, P. et al. 2012,  Proc. SPIE,  8477,  166

\bibitem[Tokovinin(2011)]{AO4ELT2}
Tokovinin, A., 2011, In: Second International Conference on Adaptive
Optics for Extremely Large Telescopes, {\url http://ao4elt2.lesia.obspm.fr}

\bibitem[Tokovinin(2013)]{AO4ELT3}  
Tokovinin,  A., 2013, In: Proceedings of the Third AO4ELT Conference. Simone
  Esposito and Luca Fini, eds. ISBN: 978-88-908876-0-4

\bibitem[Tokovinin(2014)]{Tok14}  
Tokovinin,  A. 2014, AJ, 148, 72





\end{thebibliography}
\end{document}